\documentclass[twocolumn,showpacs,preprintnumbers,amsmath,amssymb]{revtex4}

\usepackage{graphicx}
\usepackage{dcolumn}
\usepackage{bm}

\begin{document}

\preprint{APS/123-QED}

\title{New measurement of neutron capture resonances of 
  $^{209}$Bi}

\author{
\normalsize
C.~Domingo-Pardo$^{1*}$,
U.~Abbondanno$^{2}$, 
G.~Aerts$^{3}$, 
H.~\'Alvarez-Pol$^4$, 
F.~Alvarez-Velarde$^{5}$, 
S.~Andriamonje$^{3}$, 
J.~Andrzejewski$^6$, 
P.~Assimakopoulos $^7$,    
L.~Audouin $^8$, 
G.~Badurek$^{9}$
% ...
\begin{center}
\normalsize The n\_TOF Collaboration\\
\end{center}
}
\affiliation{
\mbox{$^{1}$Instituto de F{\'{\i}}sica Corpuscular, CSIC-Universidad de Valencia, Spain}\\
$^{2}$Istituto Nazionale di Fisica Nucleare, Trieste, Italy\\ 
$^3$CEA/Saclay - DSM, Gif-sur-Yvette, France\\  
$^{4}$Universidade de Santiago de Compostela, Spain\\  
$^{5}$Centro de Investigaciones Energeticas Medioambientales y Technologicas, Madrid, Spain \\  
$^{6}$University of Lodz, Lodz, Poland\\ 
$^{7}$University of Ioannina, Greece\\  
\mbox{$^{8}$Forschungszentrum Karlsruhe GmbH (FZK), Institut f\"{u}r Kernphysik, Germany}\\  
\mbox{$^9$Atominstitut der \"{O}sterreichischen Universit\"{a}ten,Technische Universit\"{a}t Wien, Austria}\\ 
\mbox{$^{10}$Centre National de la Recherche Scientifique/IN2P3 - IReS, Strasbourg, France}\\  
$^{11}$Charles University, Prague, Czech Republic\\
$^{12}$Universitat Politecnica de Catalunya, Barcelona, Spain\\  
\mbox{$^{13}$International Atomic Energy Agency, NAPC-Nuclear Data Section, Vienna, Austria}\\
$^{14}$Universidad de Sevilla, Spain\\  
$^{15}$Instituto Tecnol\'{o}gico e Nuclear(ITN), Lisbon, Portugal\\  
$^{16}$CERN, Geneva, Switzerland\\  
$^{17}$LIP - Coimbra \& Departamento de Fisica da Universidade de Coimbra, Portugal\\  
$^{18}$Istituto Nazionale di Fisica Nucleare, Bari, Italy\\  
$^{19}$University of Notre Dame, Notre Dame, USA\\
$^{20}$Centre National de la Recherche Scientifique/IN2P3 - IPN, Orsay, France\\  
$^{21}$Universit\`a degli Studi Pavia, Pavia, Italy\\  
$^{22}$Aristotle University of Thessaloniki, Greece\\  
\mbox{$^{23}$Joint Institute for Nuclear Research, Frank Laboratory of Neutron Physics, Dubna, Russia}\\  
\mbox{$^{24}$Dipartimento di Fisica, Universit\`a di Torino and Sezione INFN di Torino, Italy}\\
\mbox{$^{25}$Institute of Physics and Power Engineering, Kaluga region, Obninsk, Russia}\\  
\mbox{$^{26}$Istituto Nazionale di Fisica Nucleare(INFN), Laboratori Nazionali di Legnaro, Italy}\\
$^{27}$Centre National de la Recherche Scientifique/IN2P3 - CENBG, Bordeaux, France\\ 
$^{28}$Los Alamos National Laboratory, New Mexico, USA\\  
$^{29}$Tokyo Institute of Technology, Tokyo, Japan\\
\mbox{$^{30}$Oak Ridge National Laboratory, Physics Division, Oak Ridge, USA}\\    
$^{31}$NCSR, Athens, Greece \\
$^{32}$Japan Atomic Energy Research Institute, Tokai-mura, Japan\\
$^{33}$National Technical University of Athens, Greece\\  
\mbox{$^{34}$Institut f\"{u}r Isotopenforschung und Kernphysik, Universit\"{a}t Wien, Austria}\\
\mbox{$^{35}$P\^ole Universitaire L\'{e}onard de Vinci, Paris La D\'efense, France}\\ 
$^{36}$CEC-JRC-IRMM, Geel, Belgium\\
\mbox{$^{37}$Department of Physics and Astronomy - University of Basel, Basel, Switzerland}\\  
$^{38}$ENEA, Bologna, Italy\\  
\mbox{$^{39}$Dipartimento di Fisica, Universit\`a di Bologna, and Sezione INFN di Bologna, Italy}
}

\email{Cesar.Domingo.Pardo@cern.ch}

\date{\today}
             
\begin{abstract}
The neutron capture cross section of $^{209}$Bi has been measured at the
CERN n$\_$TOF facility by employing the pulse height weighting technique. 
Improvements over previous measurements are mainly due to an optimized
detection system, which led to a practically negligible neutron
sensitivity. Additional experimental sources of systematic error, like 
the electronic threshold in the detectors, summing of $\gamma$-rays, internal
electron conversion, and the isomeric state in bismuth, have been taken 
into account. $\gamma$-Ray absorption effects inside the sample have been 
corrected by employing a non-polynomial weighting function. Since $^{209}$Bi is 
the last stable isotope in the reaction path of the stellar $s$-process,
the Maxwellian averaged capture cross section is important for
the recycling of the reaction flow by $\alpha$ decays. In the relevant stellar
range of thermal energies between $kT=5$~keV and 8~keV our new capture rate is 
about 16\% higher than the presently accepted value used for nucleosynthesis 
calculations. At this low temperature an important part of the heavy
Pb-Bi isotopes are supposed to be synthesized by the $s$-process in the
He shells of low mass, thermally pulsing asymptotic giant branch stars. 
With the improved set of cross sections we obtain an $s$-process fraction of
19$\pm$3\% of the solar bismuth abundance, resulting in an $r$-process residual of
81$\pm$3\%.
The present ($n, \gamma$) cross section measurement is also of relevance for
the design of accelerator driven systems based on a liquid metal Pb/Bi
spallation target. 
\end{abstract}


\maketitle

\section{\label{sec:intro}Introduction}

$^{209}$Bi is the end point isotope of the $s$-process path. Its observed 
abundance has been finally understood in terms of the main $s$-process 
component, operating in the He-shell of low mass, low metallicity, thermally
pulsing asymptotic giant branch (\textsc{agb}) stars~\cite{tra99,tra01}.

\begin{figure}[h]
\includegraphics[width=0.4\textwidth]{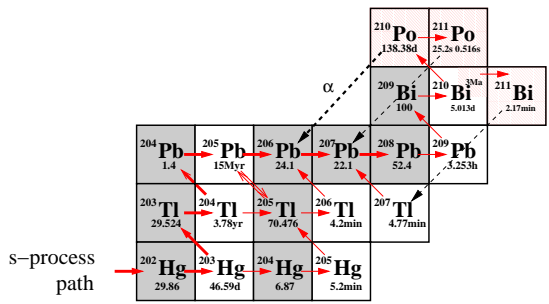}
\caption{\label{fig:EndRegion}(Color online) The $s$-process reaction network terminates at
  $^{209}$Bi. Shaded boxes designate stable
  isotopes. Decays from the $\alpha$-unstable $^{210,211}$Po isotopes are
  represented with dashed lines. Quantities below the isotopic symbols show
  natural abundances or half lifes.}
\end{figure}

However, the complex production pattern at the end of the $s$-process path
still hides important information, that can be unraveled by means of
accurate neutron capture measurements. Capture on bismuth leads to the 
ground state of $^{210}$Bi, which $\beta$-decays to $^{210}$Po that is 
$\alpha$-unstable, producing $^{206}$Pb (see Fig.~\ref{fig:EndRegion}). 
However, given the $^{210}$Po 
half-life of 138~days it can also act as a branching point by capturing 
another neutron and enhancing the abundance of $^{207}$Pb instead. 
Another contribution to $^{207}$Pb may be due to a long-lived isomer 
in $^{210}$Bi at 271~keV. Neutron capture on this isomer leads to 
$^{211}$Bi with an $\alpha$-decay half-life of 2.17 min. The two 
branchings, at $^{210}$Po and $^{210}$Bi, depend strongly on the 
stellar conditions of neutron density and temperature. Given an 
appropriate stellar model and accurate neutron capture cross sections, 
the $s$ abundances of the Pb/Bi isotopes can be more reliably determined, 
thus providing a better decomposition of the respective $r$-process
abundances and of the radiogenic contributions due to the Th/U $\alpha$
decays. 
The latter information represents a viable constraint on 
the Th/U abundances calculation and its use as cosmochronometers~\cite{cow91,kra04}.

For the discussion of these astrophysical aspects, the present status 
of the $^{209}$Bi capture data is rather unsatisfactory \cite{rat04}.
Previous experiments exhibit significant discrepancies suggesting the
existence of systematic uncertainties, which must be clearly reduced for a
more quantitative assessment of the Pb/Bi abundances. Two capture resonance studies on $^{209}$Bi 
have been reported so far, one at ORNL~\cite{mac76} and the other at 
GELINA~\cite{mut97,mut98}. In the former measurement, two strong s-wave resonances at low
neutron energy could not be measured because of a low neutron energy cutoff at 2.6~keV.
In addition, substantial corrections were needed for some of the resonances
in order to account for the neutron sensitivity of the experimental setup
employed.
The GELINA measurement was also affected by significant corrections for
neutron sensitivity. 

The $^{209}$Bi ($n, \gamma$) cross section has also a very practical 
application. Thanks to its properties of chemical inertness, high
boiling point, low neutron moderation, and large scattering cross 
section, an eutectic mixture of lead and bismuth is presently 
considered as a very appropriate material for the spallation target 
and as the coolant for accelerator driven systems (ADS).

In a first sensitivity study of the neutron cross sections of the main 
materials used in this type of hybrid reactor~\cite{rub00}, severe 
discrepancies where found between different evaluated nuclear data 
files. As a consequence, a list of isotopes was tagged for high 
priority measurements. Initially, most of the requested cross sections 
corresponded to the fuel and cladding materials, and were focused on 
the elastic, inelastic and ($n, xn$) channels. However, a recent extension 
of the same study~\cite{ado03} includes additional requests for the
capture reactions in the spallation/coolant materials lead and bismuth.
Comparing two evaluated data files, ENDF/B-VI.8 and JENDL 3.3, 
discrepancies in the neutron balance of 12.5\% were identified to 
result from the uncertainties of the ($n,\gamma$) cross section of 
$^{209}$Bi alone. Hence, the precise knowledge of the neutron capture 
cross sections of the lead and bismuth isotopes turned out to be
of key relevance for the design of an ADS suited for the transmutation 
of radioactive residues and for energy production.

Apart from its contribution to the neutron balance, neutron capture on 
$^{209}$Bi affects the radiotoxicity as well. Build-up of the 
$\alpha$-emitters $^{210}$Bi$^m$ ($t_{1/2}$ = 3$\times$10$^6$ yr) 
and $^{210}$Po ($t_{1/2}$ = 138 d) contributes to the long and short 
term radiotoxicity of the spallation target and of the coolant. These
hazards can only be evaluated on the basis of a precise ($n, \gamma$) 
cross section of bismuth and with accurate data concerning the partial
cross sections leading to $^{210}$Bi$^m$ and $^{210}$Po.

\section{\label{sec:CS measurement} Cross Section measurement}

The neutron energy differential capture cross section of $^{209}$Bi 
has been measured at the n$\_$TOF installation of CERN using the time of 
flight (TOF) technique. Since this facility has been described in detail 
elsewhere \cite{preport}, only the main features will be summarized in 
section~\ref{sec:nTOF}. The major motivation for this measurement was 
to overcome the systematic uncertainty related to the neutron scattering 
background, which arises in measurements on isotopes with a dominant
scattering channel. Sample scattered neutrons are easily captured in 
the materials surrounding the detector or in the sensitive detector
volume itself, producing $\gamma$-rays, which appear as a prompt, 
TOF-dependent background. This effect becomes particularly crucial for 
the broad $s$-wave resonances in bismuth at low neutron energies
between 0.8~keV and 12~keV, resulting in corrections as large as 50\% 
in previous measurements~\cite{mac76}.

For this reason the reaction yield of this isotope has been measured 
with a set of total energy detectors consisting of optimized C$_6$D$_6$ 
liquid scintillation detectors (section~\ref{sec:setup}). This setup 
turned out to be of pivotal importance for the efficient suppression of
the background from scattered neutrons. Other sample-related corrections 
were significantly reduced by choosing a comparably small bismuth sample 
20~mm in diameter and 6.1~mm in thickness. The cross section is 
determined relative to the 4.9~eV $^{198}$Au resonance by employing 
the saturated resonance technique~\cite{mac79}. For this purpose a gold 
sample 1~mm in thickness and 20~mm in diameter was measured in regular 
intervals. A thorough analysis of the capture data, including extensive 
Monte Carlo simulations of the necessary corrections to the experimental 
data, allowed us to determine the final cross section with a systematic 
uncertainty of better than 3\%. 

\subsection{\label{sec:nTOF}The n$\_$TOF installation}

At n$\_$TOF, neutrons are produced via spallation reactions induced by 
20~GeV protons in a lead target. Pulses of (3-7)$\times$10$^{12}$ protons 
with a duration of 6~ns (rms) hit the Pb-block with a typical repetition 
rate of 0.4 Hz. This very low duty cycle combined with the very high 
intensity in the beam pulses makes n$\_$TOF an ideal facility for TOF  
measurements on radioactive samples. The low repetition rate has allowed
to implement an advanced data acquisition system with zero dead time 
based on Flash-amplitude-to-digital converters \cite{abb05}.

The number of neutrons arriving at the sample are monitored by means 
of a 200~$\mu$g/cm$^2$ thick $^{6}$Li-foil, centered in the beam. 
The particles emitted in the $^{6}$Li($n, \alpha$)$^{3}$H reaction 
are registered with four silicon detectors surrounding the $^{6}$Li-foil 
outside of the beam. The neutron monitor~\cite{mar04} is mounted in 
an evacuated carbon fiber chamber 3 m upstream of the samples. 
The samples, which are positioned at a flight path of 185 m, are 
fixed on a sample changer, which is also made of carbon fiber.

The neutron intensity has been determined by means of two independent
measurements, performed with the Si-monitor described above and with a
calibrated fission chamber~\cite{ptb}. Two measurements were carried out 
with the latter detector, employing foils of $^{235}$U and $^{238}$U. 
Calibration of the absolute yield in ($n, \gamma$) measurements via the 
saturated resonance method requires actually only the shape of the 
neutron flux, which can be directly determined from the Si-monitor and
fission chamber measurements with an uncertainty below $\pm$2\%~\cite{dom05}.

The beam profile at the sample position has been determined
to exhibit an approximate Gaussian shape with $\sigma = 7$~mm, 
slightly off centered ($\Delta_x = 1.5$~mm) \cite{pan04}. The 
samples used in this measurement cover about 60\% of the neutron 
beam.

The excellent neutron energy resolution of the n$\_$TOF facility is 
due to the long flight path of 185~m and to the short proton pulse 
width. The resolution function was determined by means of Monte 
Carlo simulations~\cite{coc02} and has been validated experimentally 
by measuring narrow $p$- and $d$-wave capture resonances on a sample 
of $^{\textrm{\scriptsize{nat}}}$Fe.

\subsection{\label{sec:setup}Experimental setup}

A general view of the experimental setup is shown in 
Fig.~\ref{fig:setup}. The detection system used with the 
pulse height weighting technique (PHWT)~\cite{mac67} had been 
optimized with respect to the sensitivity for scattered 
neutrons, the most crucial source of background in
($n, \gamma$) studies on neutron magic nuclei, which
are characterized by particularly large scattering to
capture ratios. Scattered neutrons may be subsequently 
captured in the detectors or in nearby materials, where
they produce $\gamma$-rays, which are likely to be 
registered with high probability, thus mimicking true 
capture events in the sample. 

In an effort to reduce 
the neutron sensitivity as much as possible, special 
C$_6$D$_6$ liquid scintillation detectors~\cite{pla03} 
have been developed consisting of materials with low
neutron capture cross sections. The main features of 
these detectors are thin-walled scintillator cells 
made of carbon fiber, which are directly glued onto the 
photomultiplier tubes, thus eliminating dead materials, 
e.g. the common quartz window of the scintillator cells.
To avoid the additional material of a support, the 
detectors were hanging on thin cords fixed at the 
ceiling. Similarly, the sample changer was made of
carbon fiber as was the frame of the sample-holder.
The samples were mounted on thin Kapton foils, which 
were glued on the frame that was much larger than the
diameter of the neutron beam. In this way, the neutron
sensitivity in the critical energy range from 1 keV to 
100 keV could be reduced by factors of three to ten 
\cite{pla03}. 

The two detectors were placed at an angle of $\sim$125$^{\circ}$ 
with respect to the sample in order 
to minimize the angular distribution effects of the 
primary capture $\gamma$ rays. Also the background due 
to in-beam $\gamma$-rays scattered in the sample was
considerably reduced in this configuration ~\cite{ect03}.

\section{\label{sec:analysis}Data analysis}

In this section, the PHWT principle will be reviewed 
together with the related systematic uncertainties of 
relevance for this type of capture measurements, followed
by the determination of the weighting factors. We describe 
the calculation of accurate yield corrections and the 
analysis procedure to determine resonance parameters.

\subsection{\label{sec:PHWT}Pulse height weighting technique \\
 and systematic uncertainties}
 
The PHWT is based on two conditions, $(i)$ that the
detector efficiency $\varepsilon_{\gamma}\ll 1$ so that at 
most only one $\gamma$-ray per capture cascade is registered,
and $(ii)$ that $\varepsilon_{\gamma}$ is proportional to the 
energy of the registered $\gamma$-ray, $\varepsilon_{\gamma}
\approx \alpha E_{\gamma}$. Under these two conditions, the 
efficiency $\varepsilon_c$ for detecting a capture event, i.e.
a cascade composed of $m$ $\gamma$-rays, becomes proportional 
to the sum energy $E_c$ of that cascade. In this case one 
obtains

\begin{equation}\label{eq:1}
\varepsilon_c = 1 - \prod^{m}_{j=1}(1-\varepsilon_{\gamma j}) 
\approx \sum^m_{j=1} \varepsilon_{\gamma j} \approx \alpha E_c,
\end{equation}

which is a constant independent of the actual de-excitation 
pattern of the nucleus produced in the capture reaction. It 
is worth noting that the approximations in Eq.~\ref{eq:1} are 
the better justified the better conditions $(i)$ and $(ii)$ are
fulfilled.

The validity of Eq.~\ref{eq:1} can be challenged by several 
experimental problems though. In particular, condition $(i)$ 
is violated if more than one $\gamma$-ray of the cascade is 
registered in the same detector. Moreover, the product as
well as the sum over the $m$ $\gamma$-rays of the cascade 
is always incomplete because of the unavoidable loss of $\gamma$-rays 
due to electronic threshold applied to the detector signals, 
due to converted transitions, and due to partial population
of the isomeric state in $^{210}$Bi at 271~keV.

For all these effects, which may influence the validity of 
Eq.~\ref{eq:1}, appropriate corrections have to be determined
as is shown in the following subsections.

As far as condition $(ii)$ is concerned, the proportional 
increase of the efficiency with $\gamma$-ray energy is 
enforced by an appropriate modification of the detector energy
response distribution $R(E)$. This is achieved by application
of a pulse height dependent weighting factor $W_i$, such 
that the weighted sum of the response for a $\gamma$-ray $j$, 
$R_{ij}$, becomes proportional to its energy $E_{\gamma j}$,
\begin{equation}\label{eq:2}
\sum_i W_i R_{ij} = \alpha E_{\gamma j}. 
\end{equation}
The second approximation in Eq.~\ref{eq:1} depends directly 
on the accuracy of the calculated weights $W_i$, which can  
be tested as discussed below. 

\subsection{\label{sec:WF} Weighting factors}

It has been shown \cite{tai02,abb03} that realistic response 
functions $R_{ij}$ for mono-energetic $\gamma$-rays with 
energies $E_{\gamma j}$ can be determined by means of Monte Carlo 
simulations for any particular setup for ($n, \gamma$) 
measurements. A set of expressions of the form of 
Eq.~\ref{eq:2} can then be used to derive the weighting 
factors $W_i$.

\begin{figure}[h]
\includegraphics[width=0.2\textwidth]{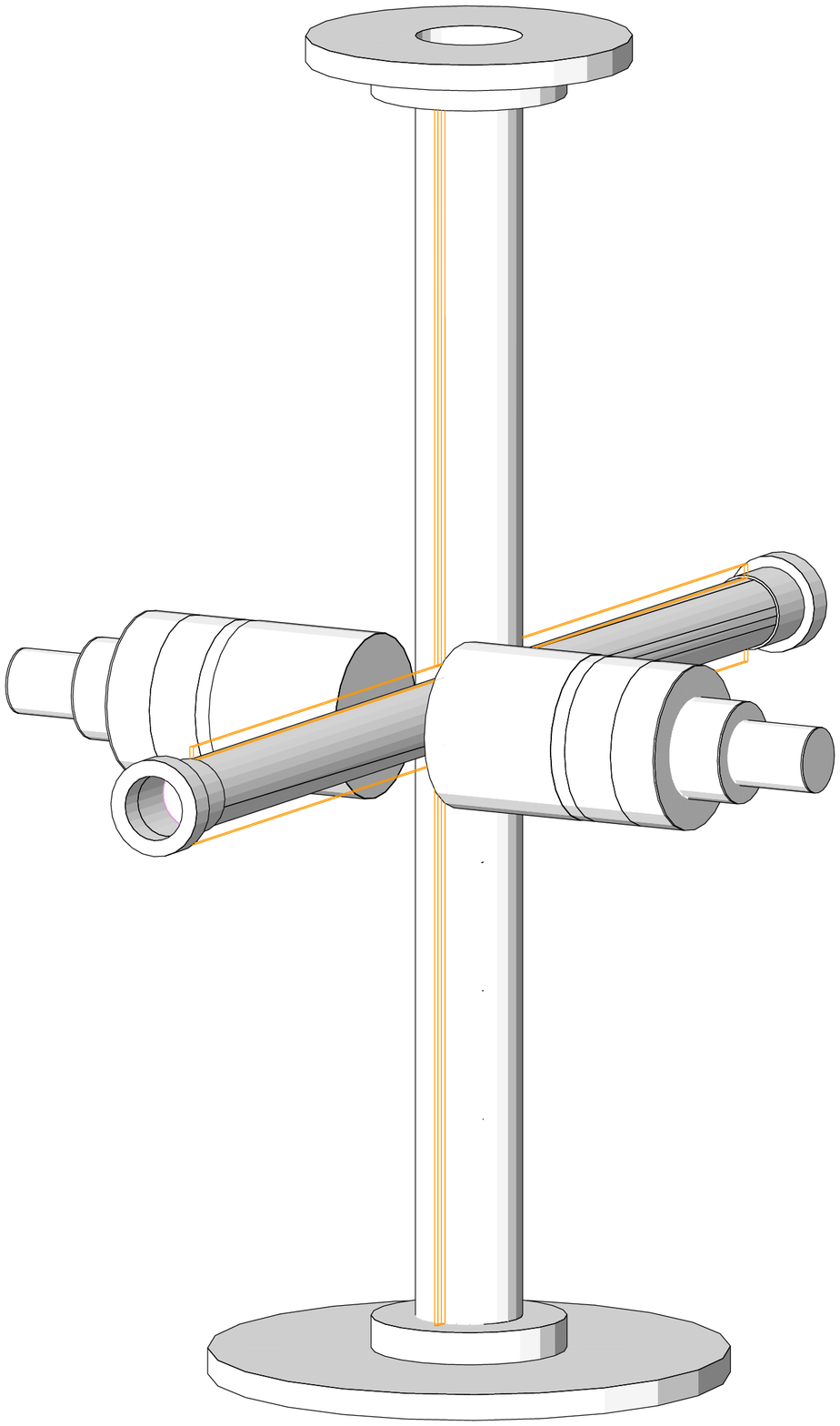}
\includegraphics[width=0.2\textwidth]{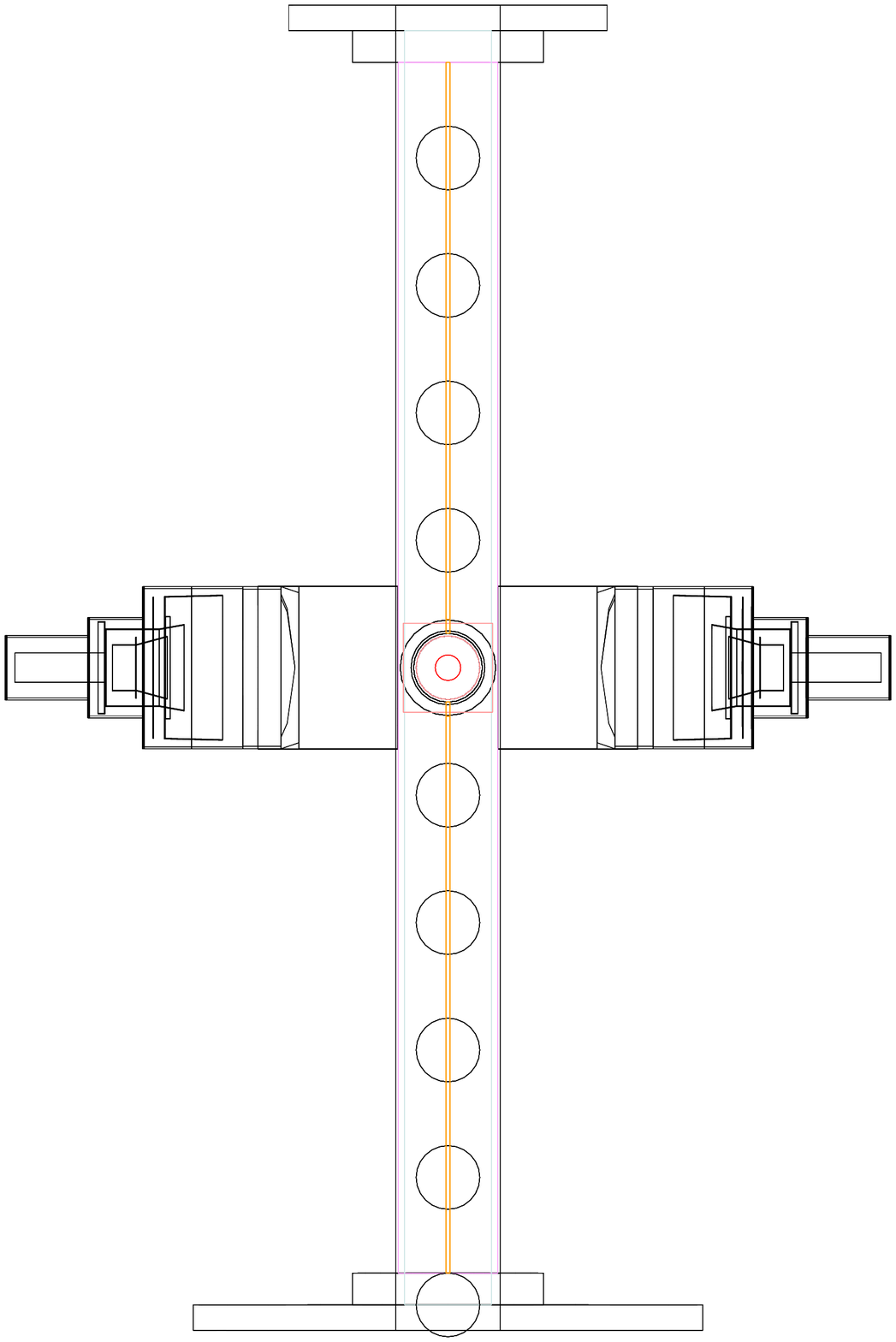}
\caption{\label{fig:setup}(Color online) The geometry of the experimental setup 
used in the  \textsc{Geant4} simulation of the weighting functions.}
\end{figure}

The detailed geometry and chemical composition of the experimental
setup has been implemented in the Monte Carlo simulation using 
\textsc{Geant4}~\cite{ago03} (see Fig.~\ref{fig:setup}). The composition of the
carbon fiber, which is by far the dominant structural material, was 
determined by an RBS analysis~\cite{dom01}, yielding C/O/N/Ca/Br 
= 2.0/0.2/0.16/0.012/0.016. Also, the capture $\gamma$-rays were 
carefully traced, assuming a radial distribution for the emission
probability inside the sample according to the neutron beam profile 
described in section~\ref{sec:nTOF}. The depth distribution was 
included as well since it changed significantly from a practically 
uniform to a surface peaked shape between weak and strong resonances,
respectively.

The weighting function (WF) of the gold calibration sample was 
obtained using the conventional approximation by a polynomial 
function of order $k=4$, $W_i = \sum_{k=0}^{k=4} a_k E^k_i$. The 
values of the coefficients $a_{k}$ were derived from a least 
squares minimization,

\begin{equation}\label{eq:3}
\mathrm{min} \sum_j \left( \sum_i W_i R_{ij} - E_{\gamma j} \right)^{2}.
\end{equation}

Because of the higher $\gamma$-ray absorption in the 6.1~mm thick 
bismuth sample the proportionality condition ($ii$) could not be 
satisfied with the polynomial WF approach. In this case an accurate, 
pointwise WF was obtained by means of a linear regularization method 
for solving the inverse problem of Eq.~\ref{eq:2}~\cite{dom05}. In 
this way the uncertainty connected with the polynomial WF has been 
reduced to 0.3\%, an improvement by an order of magnitude. 

The systematic uncertainty introduced by the WF was determined
as described in Ref.~\cite{abb03}, basically by performing Monte 
Carlo simulations of the capture cascades at a certain resonance. 
The de-excitation pattern is modeled by combining the experimentally 
known levels at low excitation energy with a statistical model 
of the nucleus to complete the cascade up to the capture energy. 
These simulated capture events can be used in order to estimate 
the uncertainty of the calculated WF.

\begin{figure}[h]
\includegraphics[width=0.23\textwidth]{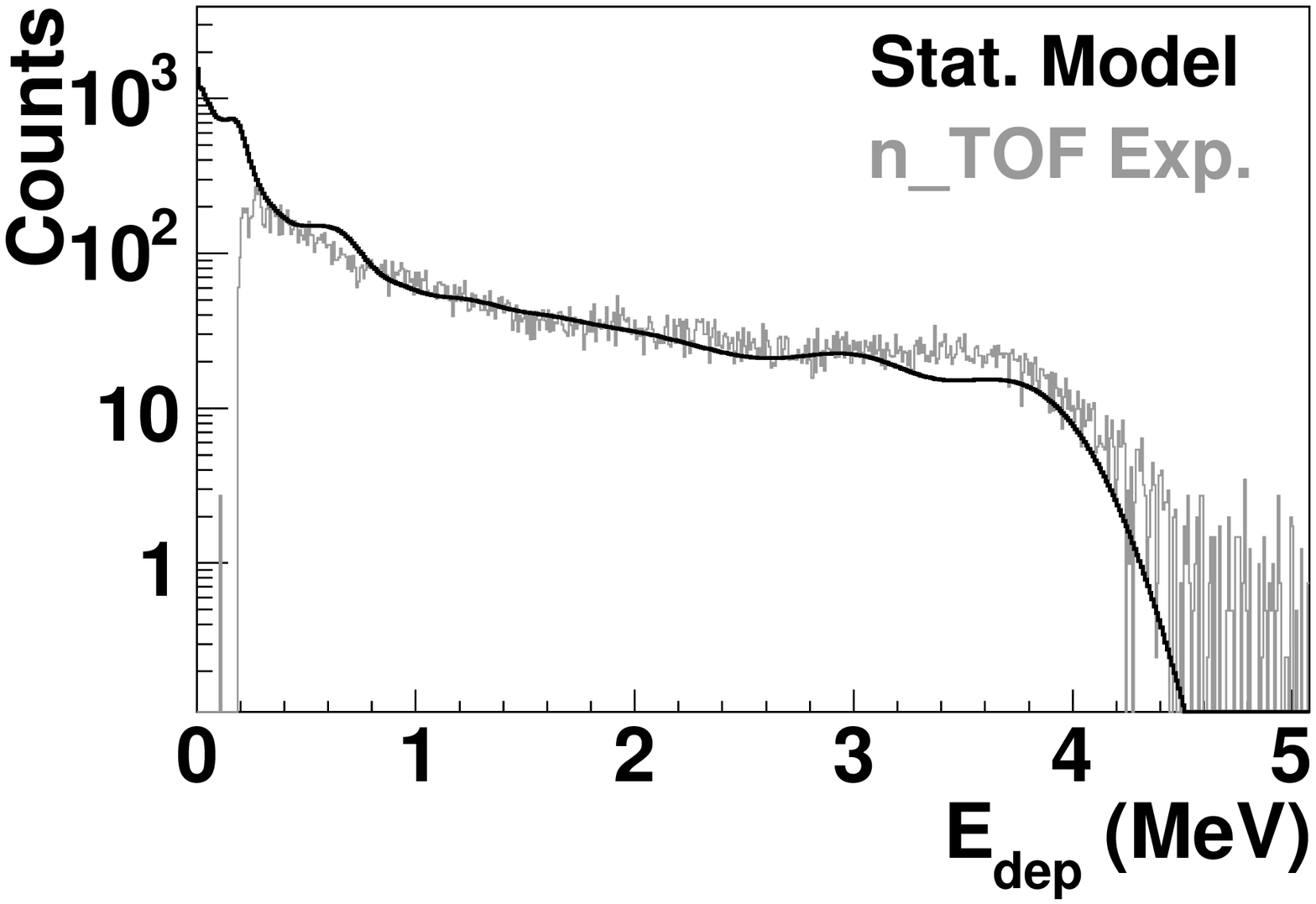}
\includegraphics[width=0.23\textwidth]{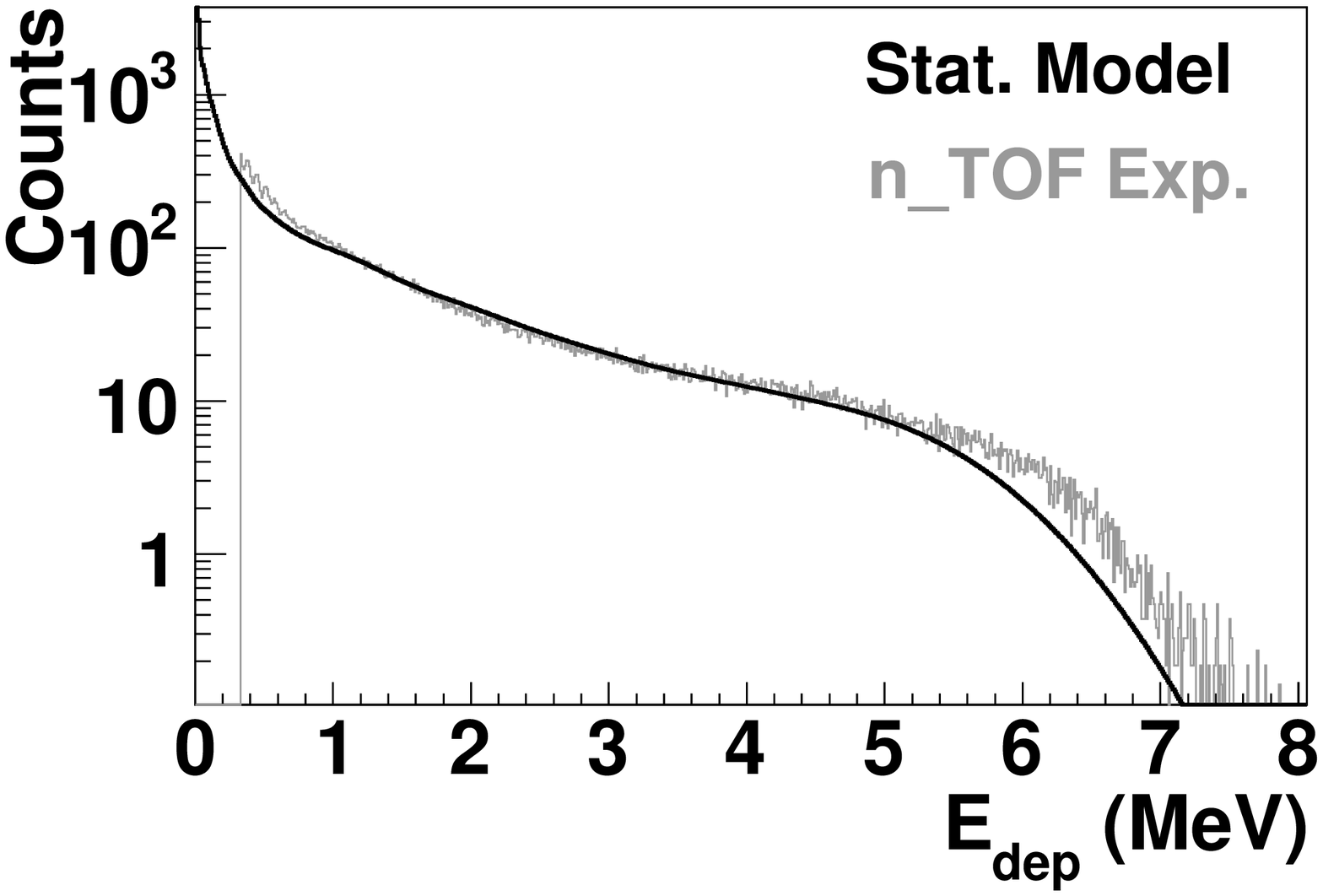}
\caption{\label{fig:BiAuCascades}Comparison of experimental (dashed) and 
simulated (solid) pulse height spectra of $^{210}$Bi (left) and 
$^{198}$Au (right) illustrating the quality of the simulations.}
\end{figure}

In order to illustrate the performance of the cascade event 
generator for the samples measured in this experiment, the 
simulated pulse height spectra $R_i^{C}$ for the 802~eV resonance 
in $^{210}$Bi and for the 4.9~eV resonance in $^{198}$Au are 
compared in Fig.~\ref{fig:BiAuCascades} with the corresponding 
experimental spectra measured at n$\_$TOF.

The spectra based on the simulated cascades show in general a good 
agreement with the experimental spectra. The small differences at 
higher energy can be ascribed to the employed statistical model and 
to a lesser extent to the uncertainties in the calibration of the 
C$_6$D$_6$ detectors. In principle, these differences could be
minimized by modifying the relative strength of M1 and E1 transitions 
in the statistical part of the cascades and/or the instrumental 
resolution. However, the effect on the weighted sum $\sum_i W_i R^{C}_i$
and, therefore, on the estimated uncertainty becomes absolutely 
negligible in both cases.

In summary, it was found that the simulated pulse height spectra can 
be employed for the determination of the uncertainty of the WF. 
By computing the weighted sum of each capture cascade with the 
calculated weighting functions, we obtain values, which deviate 
by less than 0.3\% from the corresponding capture energies. This 
means that condition ($ii$) is fulfilled within 99.7\%, leading to a
practically negligible uncertainty due to the calculated WF.

\subsection{\label{sec:threshold} Yield correction factors}

The corrections for the electronic threshold, $\gamma$-summing,
internal conversion, and the effect of the isomeric state 
(See Sec.~\ref{sec:PHWT}) have been estimated by detailed Monte Carlo 
simulations of the capture events in a similar manner as in 
Ref.~\cite{abb03}. While internal conversion and summing of 
cascade $\gamma$-rays were found to have a minor effect of 
less than 2\% on the capture yield of the bismuth sample, more 
significant corrections are required for the electronic 
threshold and for the isomeric state in $^{210}$Bi. The 
isomeric state is included in the simulation by assigning a 
null branching ratio to the level of 271~keV, so that the 
de-excitation cascade ends at that level whenever it is reached. 

The yield correction factor including all these effects can be
calculated as

\begin{equation}\label{eq:5}
f^{t,s,ce,m} = \frac{n E_c}{\sum_{i=t} W_i R^C_i},
\end{equation}

where $R^C_i$ is the total response of the detection system for $n$
simulated capture events. For a threshold $t$ of 200~keV, the yield 
correction factors calculated with this procedure for the bismuth
sample are listed in the third column of Table~\ref{tab:fcorr}.

\begin{table}[h]
\caption{\label{tab:fcorr} Yield correction factors for the bismuth 
  sample calculated for a threshold of 200~keV.}
\begin{ruledtabular}
\begin{tabular}{cccc}
$J^{\pi}$ & isom. popul.(\%) & $f^{t,s,ce,m}$ & $f^{t,s,ce,100\% m}$ \\
\hline\\
$3^{+}$   & 6                & 1.124(2)       & 1.103(2)\\
$4^{-}$   & 8                & 1.131(2)       & 1.118(2)\\
$4^{+}$   & 9                & 1.143(2)       & 1.116(2)\\
$5^{-}$   & 23               & 1.137(2)       & 1.109(2)\\
$5^{+}$   & 22               & 1.142(2)       & 1.119(2)\\
$6^{+}$   & 44               & 1.134(2)       & 1.114(2)\\
\end{tabular}
\end{ruledtabular}
\end{table}

Since the simulated spectra for resonances with same spin and 
parity are very similar, i.e. almost independent of the 
resonance energy, the correction factors could be classified 
according to spin and parity of the resonance (first column in 
Table~\ref{tab:fcorr}).

The population of the isomeric state calculated with our 
statistical model of the nucleus is shown in the second column 
of Table~\ref{tab:fcorr}. For comparison, and to illustrate the 
possible uncertainty of the calculation, the correction factor 
of the hypothetic case, where the isomer is populated with
100\% probability is shown in the last column. The differences
were found to range between 1\% and 2.5\%.

As mentioned before, the cross section is determined relative to 
the 4.9~eV resonance in $^{197}$Au. Therefore, the capture yield 
measured with the gold sample had to be corrected for experimental 
effects of threshold, $\gamma$-ray summing, and internal conversion. 
The corresponding correction factor for a threshold of 200~keV is 
$f^{t,s,ce}_{Au} = 1.046(2)$.

Hence, the final correction for each resonance is the ratio 
between the corresponding factor $f^{t,s,ce,m}$ from Table~\ref{tab:fcorr} and 
$f^{t,s,ce}_{Au}$. This ratio oscillates between 5\% and 7\%. 
It can be concluded, that the threshold, $\gamma$-summing and
internal conversion effects do not cancel out by
measuring with respect to a reference sample. Neglecting 
these corrections could lead to systematic deviations in the yield determination of
$\gtrsim$5\%.

\subsection{\label{sec:nSensitivity} Neutron sensitivity}

For a certain resonance at energy $E_{\circ}$ with capture and 
neutron widths, $\Gamma_{\gamma}$ and $\Gamma_n$, the probability 
that a signal in the C$_6$D$_6$ scintillator is caused by a neutron
scattered in the sample and eventually captured in the detector,
is given by

\begin{equation}\label{eq:6}
P^{ns} = 
\left(\frac{\varepsilon_n}{\varepsilon_{c}}\right)
\left(\frac{\Gamma_n}{\Gamma_{\gamma}}\right),
\end{equation}

where $\varepsilon_n$ denotes the probability to detect a 
$\gamma$-ray produced by sample-scattered neutrons in or near the 
detector and $\varepsilon_{c}$ the probability to register a
$\gamma$-ray from a true capture cascade. This requires a
correction of the experimental resonance yield by

\begin{equation}\label{eq:7}
f^{ns} = \frac{1}{1+P^{ns}}.
\end{equation} 

Eq.~\ref{eq:6} can be rewritten as,

\begin{equation}\label{eq:8}
P^{ns} =  \left(\frac{\varepsilon_n}{\varepsilon_{\gamma}}\right)
          \left(\frac{\varepsilon_{\gamma}}{\varepsilon_{c}}\right)
          \left(\frac{\Gamma_n}{\Gamma_{\gamma}}\right),
\end{equation}

where $\varepsilon_{\gamma}$ is the detection probability for a 
$\gamma$-ray of a given energy. For an energy $E_{\gamma}=600~$keV,  
$\varepsilon_{n}/\varepsilon_{\gamma}$ has been calculated by means
of Monte Carlo simulations for comparison with the experimentally 
determined value reported in Ref.~\cite{pla03}. 
The second factor, $\varepsilon_{\gamma}/\varepsilon_{c}$, has been 
determined by a detailed Monte Carlo simulation of the present
experimental setup. The efficiency $\varepsilon_{\gamma}$ was 
obtained for $\gamma$-rays of 600~keV, whereas $\varepsilon_c$
was determined using simulated capture cascades (see sections
\ref{sec:WF} and \ref{sec:threshold}). Finally, 
$\Gamma_n/\Gamma_{\gamma}$ is to be determined by an iterative 
procedure of correcting and fitting the value of $\Gamma_{\gamma}$ 
of the corresponding resonance.

In case of the $s$-wave resonance at 12.1~keV, which has the 
largest $\Gamma_n/\Gamma_{\gamma}$ of all analyzed bismuth 
resonances, the corresponding ratios were found to be
$\varepsilon_{\gamma}/\varepsilon_{c} = 0.446$ (this
work) and $(\varepsilon_{n}/\varepsilon_{\gamma})_{12.1~keV} 
= 1.5841 \times 10^{-5}$ \cite{pla03}. 
Because of the very low neutron sensitivity the required 
correction for the 12.1~keV resonance of 3.6$\pm$0.7\% 
is relatively small, and practically negligible for the rest 
of the $s$-wave resonances in bismuth.

The 20\% uncertainty of this correction results from neutron 
captures in the quartz window of the photomultiplier tube due to 
the uncertain silicon cross sections, which contribute a 10\% 
uncertainty to $\varepsilon_n/\varepsilon_{\gamma}$, and 
from the resonance parameters $\Gamma_n$ and $\Gamma_{\gamma}$,
which exhibit uncertainties of 8\% and 10\%, respectively.

\subsection{\label{sec:rmatrix} R-matrix analysis}

The experimental yield 

\begin{equation}
Y^{exp}=f^{t,s,ce,m}f^{ns}f^{sat}\frac{N^w}{N_n E_c},
\end{equation}

was determined by the weighted net count rate ($N^w$), 
the effective binding energy $E_c$, the integrated neutron 
flux $N_n$ (obtained from the Si-monitor and from the
shape of the neutron flux, sec.~\ref{sec:nTOF}), and the 
corrections discussed before. The factor $f^{sat}$ 
corresponds to the absolute normalization via the analysis 
of the saturated resonance at 4.9~eV in $^{198}$Au.

The yield has been analyzed with the multilevel 
R-matrix code SAMMY~\cite{lar00}. Using an iterative 
procedure based on Bayes' theorem, $Y^{exp}$ is fitted 
with a function of the type,

\begin{equation}
Y^{exp} = Y^f(E_{\circ},\Gamma_n,\Gamma_{\gamma}) + B,
\end{equation}

where $Y^f$ corresponds to a parameterization as a function 
of the resonance parameters 
$E_{\circ},\Gamma_n$ and $\Gamma_{\gamma}$
according to the Reich-Moore formalism. For all partial 
waves a channel radius of 9.6792~fm was used. While adopting 
the well known $\Gamma_n$ values from transmission
measurements, the resonance energies $E_{\circ}$ and capture 
widths $\Gamma_{\gamma}$ have been fitted to the experimental 
data. The background below the resonance is described by
the constant term $B$. In the fits with SAMMY also sample 
effects (single and double neutron scattering inside the 
sample) as well as thermal broadening are taken into account.
The resolution function of the facility (sec.~\ref{sec:nTOF}) 
has been considered in the fits using the RPI parameterization.

\begin{figure}[h]
\includegraphics[width=0.23\textwidth]{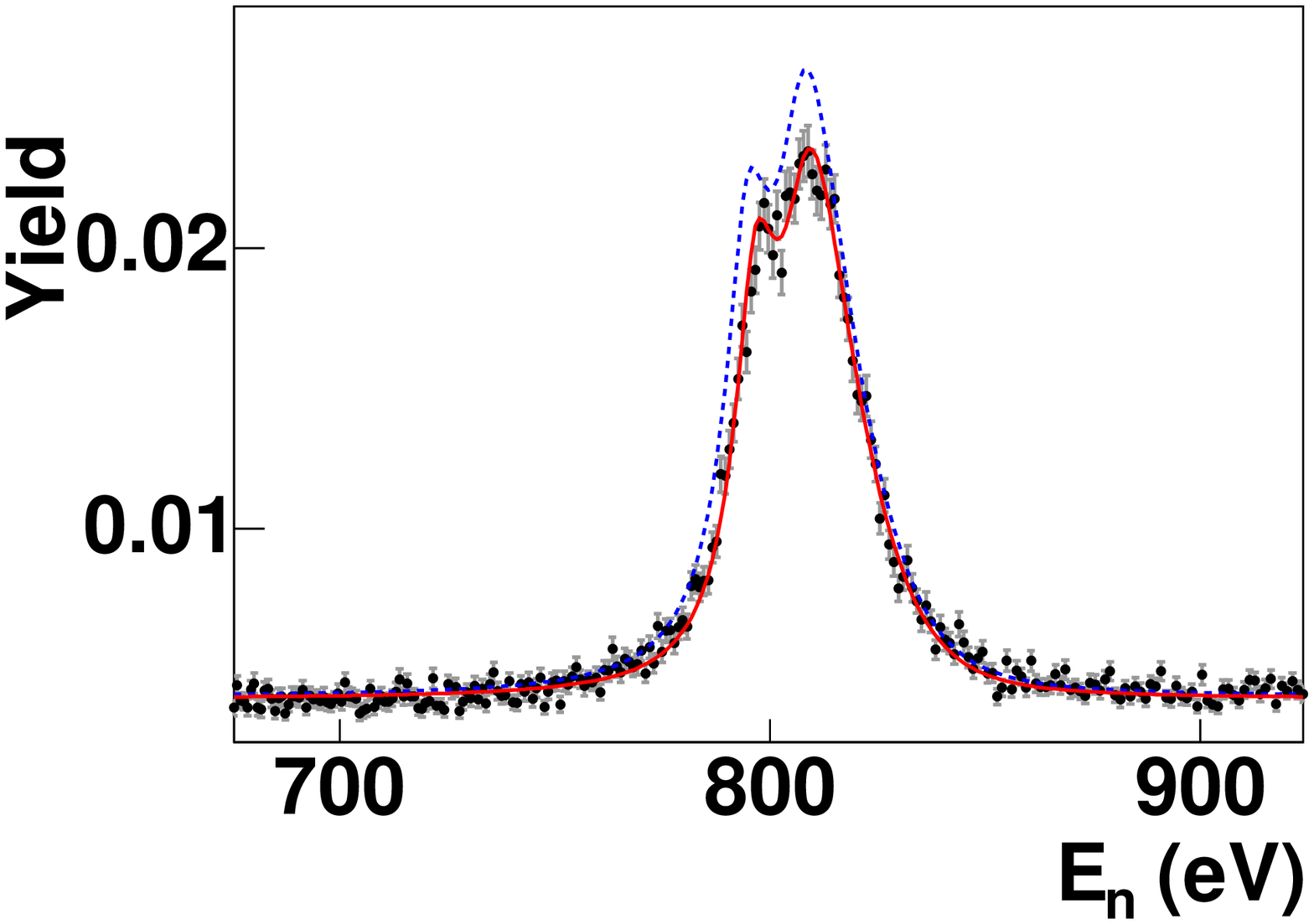}
\includegraphics[width=0.23\textwidth]{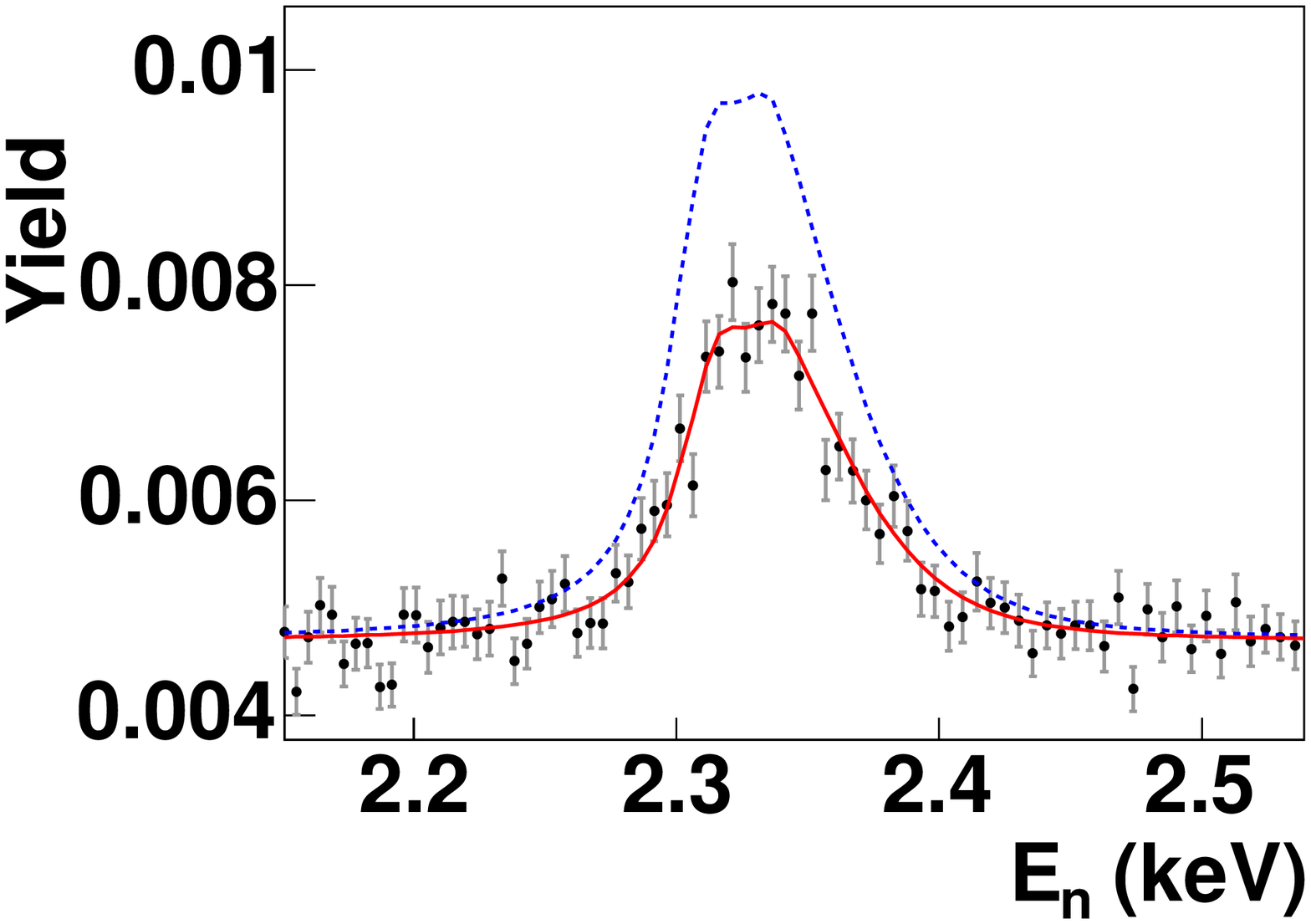}
\caption{\label{fig:fits} (Color online) R-matrix analysis of the first two 
  resonances in bismuth. The dashed line corresponds to the 
  yield calculated with the resonance parameters quoted by the 
  ENDF/B-VI.8 evaluation.}
\end{figure}

The present results are illustrated in Fig.~\ref{fig:fits} at 
the example of the first two $s$-wave resonances in $^{210}$Bi.
The comparison with the capture yields calculated with the
corresponding resonance parameters quoted in the ENDF/B-VI.8 
evaluation demonstrates the improvement due to the reduced
systematic uncertainties of the n$\_$TOF data. 

\section{\label{sec:results} Results}

In total, 21 resonances were identified in the $^{209}$Bi($n, \gamma$)
data in the energy range from 800~eV up to 23150~eV. The respective 
capture widths and areas are listed in Table~\ref{tab:bi RP and
RK}. 

\begin{table}[htbp]
\caption{\label{tab:bi RP and RK} Resonance parameters$^a$ and
  radiative kernels$^b$ for $^{209}$Bi.}
\begin{ruledtabular}
\begin{tabular}{cccccc}
E$_{\circ}$ (eV) 
        & $l$ & $J$  & $\Gamma_n$ (meV) & $\Gamma_{\gamma}$ (meV) 
		          & $g\Gamma_{\gamma}\Gamma_n/\Gamma$ (meV)\\
\hline
801.6(1)     & 0 & 5 & 4309(145)    & 33.3(12) & 18.2(6)  \\
2323.8(6)    & 0 & 4 & 17888(333)    & 26.8(17) & 12.0(8)  \\
3350.83(4)   & 1 & 5 & 87(9)        & 18.2(3)   & 9.5(2)   \\
4458.74(2)   & 1 & 5 & 173(13)      & 23.2(22) & 11.3(11)\\
5114.0(3)    & 0 & 5 & 5640(270)    & 65(2)     & 35.3(11)\\
6288.59(2)   & 1 & 4 & 116(18)  & 17.0(17) & 6.7(7)   \\
6525.0(3)    & 1 & 3 & 957(100)     & 25.3(14) &  8.6(5)  \\
9016.8(4)    & 1 & 6 & 408(77)      & 21.1(14) & 13.0(9)  \\
9159.20(7)   & 1 & 5 & 259(45)      & 21.4(21) & 10.9(11)\\ 
9718.910(1)  & 1 & 4 & 104(22)      & 74(7)     & 19.5(21)\\
9767.2(3)    & 1 & 3 & 900(114)      & 90(8)     & 28.7(26)\\
12098        &   &   &              &           & 65(4)$^{c}$\\
15649.8(1.0) & 1 & 5 & 1000     & 47(4)     & 20.2(17)\\
17440.0(1.3) & 1 & 6 & 1538(300)    & 32(3)     & 20.4(18)\\
17839.5(9)   & 1 & 5 & 464(181)    & 43(4)     & 21.7(20)\\
20870        & 1 & 5 & 954(227)    & 34.4(33) & 18.3(17)\\
21050        & 1 & 4 & 7444(778)    & 33(3)     & 14.8(13)\\
22286.0(9)   & 1 & 5 & 181(91)     & 33.6(32) & 15.1(15)\\
23149.1(1.3) & 1 & 6 & 208(154) & 25.3(25) & 14.7(15)\\
\end{tabular}
\end{ruledtabular}
$^a$Angular orbital momenta, $l$, resonance spins $J$, and
neutron widths, $\Gamma_n$, are mainly from Refs.~\cite{mug84,suk98}.\\
$^b$Uncertainties are given as 18.2(6)$\equiv$18.2$\pm$0.6.\\
$^c$This area corresponds to the sum of the areas of the broad s-wave resonance at the
indicated energy, plus two p-wave resonances at 12.092 and 12.285~keV.\\
\end{table}

Beyond 23~keV further resonances could not be observed due to 
in-beam $\gamma$-ray background~\cite{ect03}, which was the 
major limitation in this experiment. However, this background
did not affect the initial aim of improving the systematic 
uncertainty of the broad $s$-wave resonances below 100~keV.

\begin{figure}[h]
\includegraphics[width=0.45\textwidth]{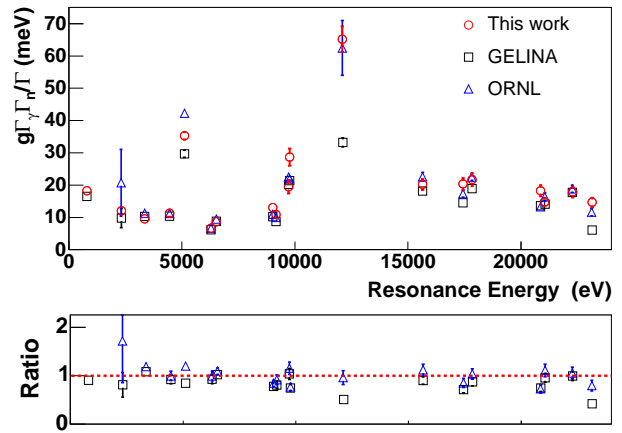}
\caption{\label{fig:bi RK} (Color online) Comparison of the radiative 
capture kernels derived from the measurements at n$\_$TOF, 
GELINA and ORNL. The capture area shown at 2.3~keV for ORNL corresponds to an
estimation reported in Ref.~\cite{mac76}. The capture area shown at $\sim$12keV, corresponds to the
sum of the three resonances in that region.}
\end{figure}

The capture areas determined at n$\_$TOF are compared in 
Fig.~\ref{fig:bi RK} with the previous measurements performed 
at ORNL~\cite{mac76} and GELINA~\cite{mut97}. 

For most of the resonances shown in Fig.~\ref{fig:bi RK}, there is a fair
agreement with the ORNL data. However, significantly higher values are
reported in the latter measurement for the two resonances at 3.3~keV and
5.1~keV, and a lower cross section for the resonance at 9.77~keV.
The GELINA data show agreement with n\_TOF in several cases, but there is an average 
trend of lower cross sections. The largest
differences in absolute values are for the resonances at 5.1~keV, 9.77~keV
and the group at 12.1~keV. For these three resonances, the ORNL data shows
opposite trend for the 5.1~keV resonance, agrees with GELINA for the 9.77~keV
and agrees with the n\_TOF data for the 12.1~keV group. 

The general systematics of better average agreement with the ORNL data and lower
average values of the GELINA data, might be related to the
weighting functions used in the respective experiments, although threshold
correction effects (see Sec.~\ref{sec:threshold}) could play a role as well. 
Systematic deviations due to the neutron sensitivity corrections applied to some particular
resonances can not be excluded also.

\subsection{\label{sec:uncertainties} Discussion of uncertainties}

The measuring technique employed in this work and the data 
analysis procedures described in Sec.~\ref{sec:analysis} 
and \ref{sec:threshold}, have been validated experimentally at
n$\_$TOF~\cite{abb03}. According to this study, the various
contributions to the systematic uncertainty of the present 
data are about 2\%. In addition, the energy dependence of 
the neutron flux has been found to exhibit an uncertainty of 
2\% as well, resulting in a total systematic uncertainty of 
better than 3\%.

In this discussion, the neutron sensitivity correction has to be
considered separately since it has been applied only to one of 
the measured resonances. Although the uncertainty of the 
correction was 20\%, the contribution to the systematic 
uncertainty of the capture kernel was only 0.7\% in this case.

\section{\label{sec:implications}Implications for the $s$-abundances 
in the Pb-Bi region}

The slightly higher cross sections for $s$-wave resonances measured 
at n$\_$TOF with respect to the values obtained at GELINA may affect
the synthesis of the Pb/Bi. This holds preferentially at lower 
stellar temperature, since the neutron sensitivity correction becomes
less relevant above 100 keV, where the relative contribution of the 
neutron scattering channel decreases.

Qualitatively, the consequence of a higher $^{209}$Bi cross section 
would reduce the survival probability and, hence, the $s$-process
abundance of this isotope, but this could be compensated to some
extent through an enhanced production through $\alpha$-recycling
via $^{206,207}$Pb.
 
In the calculation of the Maxwellian averaged cross sections (MACS), 
the n$\_$TOF resonances have been complemented between 25~keV and 
31~keV with resonances from Ref.~\cite{mut97}, and in the interval 
from 31~keV up to 60~keV with capture areas from Ref.~\cite{mug84}. 
The additional resonances represent 2\% and 7\% of the MACS at 
$kT=5$~keV and $kT=8$~keV, respectively. At lower stellar temperature, 
comparison of the results in Table~\ref{tab:bi MACS} with the MACS 
obtained at GELINA~\cite{mut98} shows that the new values are 16\% 
larger.

\begin{table}[h]
\caption{\label{tab:bi MACS}Maxwellian averaged cross 
sections  of $^{209}$Bi compared to recent previous data.}
\begin{ruledtabular}
\begin{tabular}{ccc}
$kT$  & This work   & Mutti et al.\cite{mut98} \\
(keV) &   (mb)          & (mb)  \\
\hline
5     &  13.05(84)&  11.25(58) \\
8     &  8.62(54) &  7.48(44)  \\
20    &  3.41(50) &  3.34(42) \\
25    &  2.89(50) &  2.85(45)  \\
\end{tabular}
\end{ruledtabular}
\end{table}

Beyond $kT=20$~keV, the MACS is strongly influenced by the average 
capture cross section above 80~keV, which has been determined 
experimentally in Ref.~\cite{mut98}. For this reason, the differences 
obtained in the strong s-wave resonances at lower neutron energy 
do not affect substantially the MACS at 20-25~keV.

The MACS given in Table~\ref{tab:bi MACS} refer to the lower
temperature regime of the common stellar $s$-process site
associated with thermally pulsing low mass \textsc{agb} stars
\cite{GAB98}. 
According to the Galactic chemical evolution (GCE) model described in Refs.~\cite{tra99,tra01},
the $s$-process abundances of $^{208}$Pb and $^{209}$Bi are essentially
produced in stars of low metallicity. In this model about 
95\% of the neutron exposure is due to the $^{13}$C($\alpha, 
n$)$^{16}$O reaction, which operates during the interpulse phase 
between He shell flashes at temperatures around $\sim 10^8$~K,
corresponding to a thermal energy of $kT \approx8$~keV. 

The additional neutron irradiation provided by the
$^{22}$Ne($\alpha, n$)$^{25}$Mg reaction at the higher thermal
energy of $kT = 23$ keV during the He shell flash is rather
weak. Hence, the abundances of isotopes with small cross sections
are still dominated by what has been produced at the lower
temperature of the interpulse phase. Therefore, the 
low temperature part of the MACS is important for this 
stellar model. With respect to the solar abundance tables of Anders and
Grevesse~\cite{and89}, the GCE model~\cite{tra01} provides $s$-process
abundances of $^{206}$Pb, $^{207}$Pb and $^{209}$Bi at the epoch of the solar
system formation of 62\%, 79\%, and 19\%, respectively.

According to the sensitivity study reported in Ref.~\cite{rat04}, 
the present result (see table~\ref{tab:bi MACS}) is not expected to affect the abundances of 
$^{206,207}$Pb by $\alpha$-recycling from $A=210$. This was confirmed by a model calculation  for a
thermally pulsing \textsc{agb} star with $M = 3 M_{\odot}$ and 
a metallicity $[\mathrm{Fe/H}]=-1.3$, which yields a negligible 
difference for the abundances of $^{206}$Pb and $^{207}$Pb. The $s$-process abundance of $^{209}$Bi
itself was found to decrease only slightly from 19\% to 18.7\%.

To evaluate the uncertainty on the solar $^{209}$Bi $s$-process fraction, we
have to consider first the uncertainty in the cross section of this isotope
in the two temperature regimes of the stellar model.
The main contribution due to the uncertainty of the bismuth cross section is
dominated by the conditions during the He shell flash. At the higher temperature
of the $^{22}$Ne($\alpha$,n)$^{25}$Mg source ($kT=23$~keV) the reaction
flow via $^{208}$Pb($n, \gamma$)$^{209}$Bi is strongly enhanced over the
situation during the $^{13}$C($\alpha$,n)$^{16}$O phase. This reflects the
increase of the $^{208}$Pb cross section with stellar
temperature~\cite{bee97}.
The  net effect of the uncertainty on the $^{209}$Bi cross section to its solar $s$-process
abundance is found to be 6\%.

In the evaluation of the final $s$-process abundance of bismuth, three
aditional uncertainties have to be considered:
(i) the effect of the uncertainty in the 
$^{208}$Pb cross section, which directly affects the $s$ production of
$^{209}$Bi by 6.5\% (see Table~VII in Ref.~\cite{rat04}), 
(ii) the 7-8\% uncertainty of the solar bismuth abundance 
(Refs.~\cite{lod03,and89}), and 
(iii) a further 10\% for the uncertainties
related to the $s$-process model and for modeling GCE~\cite{tra01}.
In summary the $s$-process contribution to solar bismuth is obtained as
19$\pm$3\%, corresponding to an $r$-process residual of 81$\pm$3\%.

This result is in agreement with the $r$-process calculation of Ref.~\cite{cow99} using the waiting
point approximation and improved mass formulae, which yield an $r$-process
contribution between 71\% and 90\%. A similar result is reported in
Ref.~\cite{kra04} where the calculations give 77\% and 92\% $r$-process
contribution depending on the initial seed composition. 

\section{\label{sec:other implications}Estimation of the thermal 
capture cross section}

The thermal neutron capture cross section of $^{209}$Bi can be 
expressed by the sum of the tails of all Breit-Wigner
resonances,

\begin{equation}\label{eq:sigma_th simp swave}
\sigma^{th}_{\gamma} \approx 4.099\times 10^6 
\left(\frac{A+1}{A}\right)^2\sum^{N_{\circ}}_{i=1} 
g^i_{n}\frac{\Gamma^i_{n}\Gamma^i_{\gamma}}{(E^i_{\circ})^{2}}.
\end{equation}

The result obtained with the resonance parameters from the present
measurement (see Table~\ref{tab:bi RP and RK}), is in agreement with that
obtained from the GELINA measurement, as shown in Table~\ref{tab:sigma
  thermal}. However, both values are around 40\% smaller than the accepted
value of 33.8(5)~mb~\cite{mug84}, which has been measured with the pile
oscillator method. The direct capture process cannot account for this
discrepancy. In fact, a preliminary estimate of this component leads to a
negligible contribution to the thermal cross section value~\cite{Arb04}.

\begin{table}[h]
\caption{\label{tab:sigma thermal}
Comparison of the thermal neutron capture cross section from 
different sources.}

\begin{ruledtabular}
\begin{tabular}{llcc}
 &  & Ref. & $\sigma^{th}_{\gamma}$ (mb)\\
\hline
In-pile measurement   &           & \cite{mug84} &  33.8(5)   \\
From resonance parameters &           &&            \\
                          & This work && 23.6(9)  \\
                          & GELINA    &\cite{mut97}& 24.6(9)  \\
                          &           &&            \\
                          & ENDF      && 32.51      \\
                          & JENDL     && 32.51      \\
\end{tabular}
\end{ruledtabular}

\end{table}

The resonance parameters of the ENDF and JENDL libraries have been
adjusted to reproduce 
the in-pile $\sigma^{th}_{\gamma}$ measurement by means of an expression like
Eq.~\ref{eq:sigma_th simp swave}. With the improved, present
data for the $s$-wave resonances this inconsistency could
be removed by introducing subthreshold resonances in the 
evaluated data files.

\section{\label{sec:summary} Summary}
At the CERN n$\_$TOF facility the time of flight method has 
been employed with the pulse height weighting technique in 
order to determine the neutron capture cross section of bismuth 
in the resolved resonance region.

The main improvement with respect to previous measurements 
is due to an optimized detection setup, by which the crucial 
neutron sensitivity could be considerably reduced. All
remaining sources of systematic uncertainties have been 
thoroughly treated by detailed Monte Carlo simulations.

Resonance energies ($E_{\circ}$), widths ($\Gamma_{\gamma}$),
and capture areas have been determined by an R-matrix analysis
covering the energy range from 0.8~keV to 23~keV. 

The results show larger capture areas for the $s$-wave 
resonances, yielding a 16\% enhancement of the stellar neutron 
capture rate at thermal energies between 5~keV and 8~keV 
compared with recent data~\cite{mut98}. 

The new cross section results for $^{209}$Bi, combined with recent improvements in the cross
section of $^{208}$Pb~\cite{rat04}, yield a solar $s$-process abundance of
19(3)\% for bismuth. The resulting $r$-process residual of 81(3)\% represents
a reliable constraint for $r$-process calculations~\cite{cow99}.

\begin{acknowledgments}
We acknowledge the help of G. Arbanas (ORNL) in providing the
direct capture contribution to the thermal cross section. 
R.G. and F.K. appreciate the opportunity to discuss this paper at the Aspen Summer School
organized in 2005 by R.~Reifarth and F.~Herwig.
This work was supported by the European Commission (FIKW-CT-2000-00107), 
by the Spanish Ministry of Science and Technology (FPA2001-0144-C05), 
and partly by the Italian MIUR-FIRB grant "The astrophysical origin 
of the heavy elements beyond Fe". It is part of the Ph.D. thesis of 
C.~D., who acknowledges financial support from Consejo 
Superior de Investigaciones Cient\'{\i}ficas.
\end{acknowledgments}

\newpage 
\bibliography{article209Bi}

\end{document}